\documentclass[12pt,preprint]{aastex}
\usepackage{natbib}
\usepackage{graphicx}
\shorttitle{Shaping the relation between the supermassive black hole and 
the velocity dispersion}
\begin{document}
\title{Shaping the relation between the mass of supermassive 
black holes and the velocity dispersion of galactic bulges}
\author{M.~H.~Chan}
\affil{Department of Physics and Institute of Theoretical Physics,
\\ The Chinese University of Hong Kong,
\\  Shatin, New Territories, Hong Kong, China}
\email{mhchan@phy.cuhk.edu.hk}

\begin{abstract}
I use the fact that the radiation emitted by the accretion disk of
supermassive black hole can heat up the surrounding gas in the protogalaxy 
to achieve hydrostatic equilibrium during the galaxy formation. The 
correlation between the black hole mass $M_{BH}$ and velocity dispersion 
$\sigma$ thus naturally arises. The result generally agrees 
with empirical fittings from observational data, even with 
$M_{BH} \le 10^6M_{\odot}$. This model provides a clear picture on how the 
properties of the galactic supermassive black holes are connected with the 
kinetic properties of the galactic bulges. 
\end{abstract} 

\section{Introduction}
In the past decade, some observations have led to some tight relations 
between the central supermassive blackhole (SMBH) masses $M_{BH}$ and 
velocity dispersions $\sigma$ in the bulges of galaxies. These relations 
can be summarized as $\log(M_{BH}/M_{\odot})= \beta \log(\sigma/200~ \rm 
km~s^{-1})+ \alpha$. For $10^6M_{\odot} \le M_{BH} \le 10^9M_{\odot}$, the 
values of $\alpha$ and $\beta$ have been 
estimated several times in the past 12 years: originally 
($\alpha,\beta$)=($8.08 \pm 0.08$, $3.75 \pm 0.3$) \citep{Gebhardt} and 
($8.14 \pm 1.3$, $4.80 \pm 0.54$) 
\citep{Ferrarese}, then ($8.13 \pm 0.06$, $4.02 \pm 0.32$) 
\citep{Tremaine}, and more recently ($8.28 \pm 0.05$, $4.06 \pm 0.28$) 
\citep{Jian}, ($8.12 \pm 0.08$, $4.24 \pm 0.41$) \citep{Gultekin}, 
($8.29 \pm 0.06$, $5.12 \pm 0.36$) \citep{McConnell} and ($8.13 \pm 0.05$, 
$5.13 \pm 0.34$) 
\citep{Graham}. Generally speaking, empirical fittings show that $\alpha 
\approx 8$ and $\beta \approx 4-5$. These relations correspond to all 
morphological type galaxies. One can separate the fittings into different 
groups such as the early-type and late-type or elliptical and spiral. For 
example, \citet{McConnell} obtain $(\alpha,\beta)=(8.38,4.53)$ and 
$(7.97,4.58)$ for the early-type and late-type galaxies respectively if 
they are fitted separately. The 
slopes are shallower than the combined one ($\beta=5.12$). Moreover, the 
slope $\beta$ and the scatter of the 
$M_{BH}-\sigma$ relation are still subject to debate, particularly at the 
low mass ends. Recently, \citet{Xiao} obtain a new $M_{BH}-\sigma$ 
relation with low BH masses (below $2 \times 10^6M_{\odot}$). They find a 
zero point $\alpha=7.68 \pm 0.08$ and slope $\beta=3.32 \pm 0.22$, which 
indicate $\beta$ may be smaller for lower BH masses. Also, 
\citet{Wyithe} obtained 
a better fit by using a log-quadratic form $\log(M_{BH}/M_{\odot})= \alpha 
+ \beta \log( \sigma/200~ \rm km~s^{-1})+ \gamma' [ \log( \sigma/200~ \rm 
km~s^{-1})]^2$ with $\alpha=8.05 \pm 0.06$, $\beta=4.2 \pm 0.37$ and 
$\gamma'=1.6 \pm 1.3$. Therefore, it is reasonable to doubt that the 
relation is not simply given by $M_{BH} \propto \sigma^{\beta}$ with 
$\beta$ just a constant for all galaxies.

The $M_{BH}-\sigma$ relation has been derived by recent 
theoretical models 
\citep{Rees,Adams,MacMillan,Robertson,Murray,King,McLaughlin,Power,Nayakshin}. 
However, these 
models contain various assumptions and fail to explain the relations 
in the small SMBH mass regime ($\beta 
\approx 3.3$). In this article, I present 
a model to get an exact $M_{BH}-\sigma$ relation, which can explain the 
parameters $\alpha$ and $\beta$ in the empirical fitting in both small 
SMBH regime and apply to different types of galaxies. I use the fact that 
the strong radiation of the accretion disk of a SMBH can heat up the 
surrounding gas so that hydrostatic equilibrium of the latter 
is maintained. The cooling of the surrounding gas is mainly given by 
recombination, bremsstrahlung radiation and the adiabatic expansion of 
the gas. Without any other 
assumptions, the exact $M_{BH}-\sigma$ relation is naturally obtained. In 
the following, I will first present the details of the model. Then I 
will fit our model with the data of $\sigma$ and $M_{BH}$ from 198 
galaxies and show that it generally agrees with the empirical 
fitting.

\section{The Accretion model of supermassive black hole and the 
$M_{BH}-\sigma$ relation}
It is commonly believed that all SMBHs accompany with accretion disks to 
emit high energy radiation during their formation. The luminosity of 
the disk is mainly come from the rest mass energy 
of the mass accretion. The accretion luminosity can be expressed 
as 
\begin{equation}
L_{\rm disk}=\eta \dot{M}c^2,
\end{equation}
where $0.05 \le \eta \le 0.4$ is the efficiency of the process. Let 
$f_{Ed}=L_{\rm 
disk}/L_{Ed}$, where $L_{Ed}=1.5 \times 10^{38}(M_{BH}/M_{\odot})$ erg 
s$^{-1}$ is the Eddington limit of accretion, we have
\begin{equation}
L_{\rm disk}=1.5 \times 10^{38}f_{Ed} 
\left(\frac{M_{BH}}{M_{\odot}} \right)~{\rm erg~s^{-1}}.
\end{equation} 

The accretion disk of SMBH provides a large number of photons to heat up 
the surrounding gas in the protogalaxy during the galaxy formation. Assume 
that the power is mainly transmitted to the protogalaxy within a 
scale radius $R$ through radiation by compton scattering and 
photoionization. The optical depth of the gas is $\tau=n 
\sigma_{ph}R \ll 1$, where $n$ is the number density of 
the hot gas and $\sigma_{ph}$ is the effective cross section of the 
interaction of photons and hot gas particles, which is closed to the 
Thomson cross section $\sigma_{Thom}$ for zero metallicity. Therefore, the 
total power 
that can be transmitted to the protogalaxy is just $L_{\rm disk} \tau$. In 
equilibrium, the heating rate is 
equal to the cooling rate by bremsstrahlung radiation $\Lambda_B$, 
recombination $\Lambda_R$, and adiabatic expansion $\Lambda_a$ 
\citep{Katz}:
\begin{equation}
L_{\rm disk} \tau=\Lambda_{B0}n^2T^{1/2}V+ \Lambda_{R0}n^2T^{0.3} 
\left(1+ \frac{T}{10^6~\rm K} \right)^{-1}V+ p \frac{dV}{dt},
\end{equation}
where $\Lambda_{B0}=1.4 \times 10^{-27}$ erg cm$^3$ s$^{-1}$, 
$\Lambda_{R0}=3.5 \times 10^{-26}$ erg cm$^3$ s$^{-1}$, $T$, $p$, $V$, 
are the temperature, pressure and volume of the gas within $R$, 
respectively. The term $pdV/dt$ can be written as $pdV/dt \approx 
pV^{2/3}(\gamma kT/m_g)^{1/2}$ and $p=nkT$ \citep{Muno,Chan}, where 
$\gamma \approx 
5/3$ is the adiabatic index and $m_g$ is the mean mass of a gas 
particle. The Virial relation of the effective total mass of hot gas 
$M_g$ and $T$ within $R$ is given by \citep{Sarazin}
\begin{equation}
kT=f_1 \frac{GM_gm_g}{3R},
\end{equation}
where $f_1$ is the virial 
constant. After the galactic bulge is formed, assuming spherical symmetry 
and by Virial theorem again, one can get
\begin{equation}
\sigma^2=f_2 \frac{GM_g}{R},
\end{equation}
where $f_2$ is another virial constant. From Eqs.~(3), (4) and (5), and 
assuming $nV=M_g/m_g$, we get
\begin{equation}
L_{\rm disk}=L_1 \sigma^3+L_{21} \sigma^{2.6}(1+L_{22} \sigma^2)^{-1}+L_3 
\tau^{-1} \sigma^5,
\end{equation}
where
\begin{equation}
L_1=\frac{\Lambda_{B0}f_1^{1/2}}{3^{1/2}m_g^{1/2}f_2^{3/2}Gk^{1/2} 
\sigma_{ph}},  
\end{equation} 
\begin{equation}
L_{21}=\frac{\Lambda_{R0}f_1^{0.3}}{3^{0.3}m_g^{0.7}f_2^{1.3}Gk^{0.3} 
\sigma_{ph}}, 
\end{equation}
\begin{equation}
L_{22}=\frac{f_1m_g}{3 \times 10^6f_2k}
\end{equation}
\begin{equation}
L_3=\frac{2.1f_1^{3/2}}{f_2^{5/2}G}. 
\end{equation}
Take $f_1 \approx 1$ (isothermal distribution) and $f_2 \approx 1/5$ 
\citep{Cappellari}, and combine with Eq.~(2), we have
\begin{equation}
\frac{M_{BH}}{10^8M_{\odot}}=f_{Ed}^{-1} \left[6.2 \sigma_{200}^3+6.3 
\sigma_{200}^{2.6}(1+10.4 
\sigma_{200}^2)^{-1}+0.37 \tau^{-1} \sigma_{200}^5 \right],
\end{equation}
where $\sigma_{200}= \sigma/200~\rm km~s^{-1}$. Therefore, assuming 
$\tau \sim 0.005$, the last term dominates for $\sigma \ge 60$ 
km~s$^{-1}$, which agrees with the observed range 
of $\beta$. The fitting 
parameters $f_{Ed}$ and $\tau$ are mainly controlled by the empirical 
fitting parameters $\alpha$ and $\beta$ of the observational data 
respectively. In 
Fig.~1, we get an empirical fit by using the data obtained from 
\citet{Greene,Xiao,McConnell}. The effective cross section due 
to metallicity may contribute to a factor of 2-3 in Eq.~(11). By using the 
cross sections of some major metals (carbon, nitrogen, oxygen, silicon) 
calculated from \citet{Daltabuit} and assuming metallicity of a
protogalaxy is about $10^{-3}$ solar metallicity \citep{Jappsen}, the 
effective ccross section is $2 \times 10^{-24}$ cm$^{-2}$, which is about 
$3 \sigma_{Thom}$. Also, the estimation of the $M_{BH}$ is not 
too reliable for $M_{BH} \le 10^6M_{\odot}$. Therefore, the fitting 
parameters are just an order of magnitude estimation. In Fig.~1, the 
functional form of Eq.~(11) 
generally matches the observational data. The best fitted 
parameters are $f_{Ed}=50$ and $\tau=0.005$, with 4.9\% rms 
error. By fitting with the form $\log(M_{BH}/M_{\odot})=\beta 
\log \sigma_{200}+ \alpha$, we can get $(\alpha, \beta)=(8.2,4.5)$ with 
5.1\% rms error. Therefore, two functional forms can fit the data equally 
well. However, if we neglect the first two terms in the right hand side of 
Eq.~(11), the best fitted line is $M_{BH}/10^8M_{\odot}=4.5 
\sigma_{200}^5$ with 12\% rms error. As mentioned above, the 
first two terms are significant when $M_{BH}$ or $\sigma$ is small. 
Therefore, the effective slope of the $\log M_{BH}-\log \sigma$ relation 
is shallower ($\beta<5$). That means the 
effect of cooling by recombination and bremsstrahlung radiation should be 
considered especially in galaxies with low velocity dipersion.  

Our result is consistent with the recent observations which indicate that 
many supermassive black holes may involve a long period 
of moderate super-Eddington accretion ($f_{Ed} \sim 10$) during their 
formation \citep{Kawaguchi,Brian,Wang}. On the 
other hand, the central number density in the Milky Way is about 
$0.1-0.5$ cm$^{-1}$ \citep{Muno}, which corresponds to $\tau \sim 
0.001-0.005$ in the bulge. The best fitted $\tau$ is also consistent with 
the observational data.

\section{Discussion}
In this article, I present a new model to explain the $M_{BH}-\sigma$ 
relation in galaxies. The $M_{BH}-\sigma$ relation is not simply a 
power-law form $M_{BH} \propto
\sigma^{\beta}$, but partially depends on $\sigma^{3}$ and $\sigma^5$. 
This exact form of the $M_{BH}-\sigma$ relation agrees with the recent
observational data, especially in the small $M_{BH}$ 
regime. In this model, we can obtain $3 \le \beta \le 5$ and $\alpha 
\approx 8$, which is consistent with the observational 
data of the grouped galaxies (small SMBH: $\alpha \approx 7.7$, $\beta 
\approx 3.3$; early-type: $\alpha \approx 8.4$, $\beta \approx 4.5$; 
late-type: $\alpha \approx 8.0$, $\beta \approx 4.6$) 
\citep{Xiao,McConnell}. In general, this model suggests that larger $\tau$ 
and $f_{Ed}$ result in smaller 
slope $\beta$ and normalization constant $\alpha$ in the relation. Thus, 
if similar galaxies have similar bulge structure 
and accretion disks, then the $M_{BH}-\sigma$ relation of this particular 
type may be tighter. It generally agrees with the observation that the 
$M_{BH}-\sigma$ relation in elliptical galaxies only is less scattered 
\citep{Graham}. 

In this model, the evolution pattern of the supermassive black hole does 
not affect the function form of the relation. The only physics here is the 
energy balance of the gas between the heating by the radiation from 
accretion disk and the cooling by the free-free emission, recombination 
and the adiabatic expansion of the gas particles. If the black hole is 
still significantly accreting, the energy given out would be balanced by 
the cooling of gas, which gives the Eq.~(11). When the black hole's 
activity is switched off, the relation between the kinematic properties of 
the bulge and the $M_{BH}$ has already been established, which remains 
unchanged in Eq.~(11). Therefore, the exact relation between $M_{BH}$ and 
$\sigma$ can definitely apply in both active and non-active galaxies. 

All the parameters obtained ($f_{Ed} \sim 10$ and $\tau 
\sim 0.001$) are consistent with the theoretical estimation 
and observation. Generally, our result supports the moderate 
super-Eddington accretion during the SMBH formation. The variations 
of $f_{Ed}$ and $\tau$ within the groups of galaxies may result in the 
observed 
scatter in the $M_{BH}-\sigma$ fittings. All the above results arose 
from existing natural physical laws without any extra assumptions. This 
model provides a clear picture on how the 
properties of the galactic supermassive black holes are connected with the 
kinetic properties of the galactic bulges. 

\section{Acknowledgement}
I am grateful to the referee for helpful comments on the manuscript.

\vskip 10mm

\begin{figure*}
\vskip5mm
 \includegraphics[width=140mm]{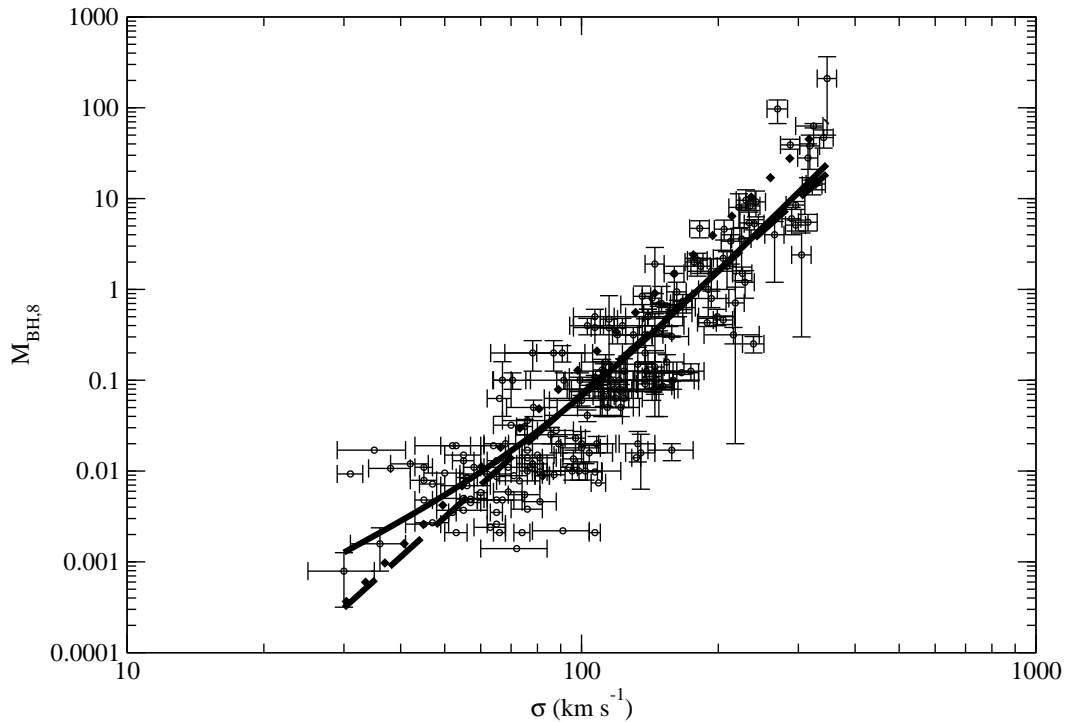}
 \caption{$M_{BH,8}$ versus $\sigma$ for 198 galaxies obtained from 
\citet{Greene,Xiao,McConnell}, where $M_{BH,8}=M_{BH}/10^8M_{\odot}$. The 
solid line is generated from Eq.~(11) with $f_{Ed}=50$ and $\tau=0.005$. 
The dashed line is in the form $\log (M_{BH}/M_{\odot})=\alpha+\beta 
\log(\sigma_{200})$ with $\alpha=8.2$ and $\beta=4.5$. The dotted line is 
$M_{BH,8}=4.5 \sigma_{200}^5$.} 
\end{figure*}

\end{document}